\documentclass[twoside,twocolumn,9pt]{article}
\usepackage{extsizes}
\usepackage[super,sort&compress,comma]{natbib}
\usepackage[version=3]{mhchem}
\usepackage[left=1.5cm, right=1.5cm, top=1.785cm, bottom=2.0cm]{geometry}
\usepackage{balance}
\usepackage{widetext}
\usepackage{times,mathptmx}
\usepackage{sectsty}
\usepackage{graphicx}
\usepackage{lastpage}
\usepackage[format=plain,justification=justified,singlelinecheck=false,font={stretch=1.125,small,sf},labelfont=bf,labelsep=space]{caption}
\usepackage{float}
\usepackage{fancyhdr}
\usepackage{fnpos}
\usepackage[english]{babel}
\usepackage{array}
\usepackage{droidsans}
\usepackage{charter}
\usepackage[T1]{fontenc}
\usepackage[usenames,dvipsnames]{xcolor}
\usepackage{setspace}
\usepackage[compact]{titlesec}
\usepackage[titletoc,toc,title]{appendix}

\usepackage{epstopdf}

\definecolor{cream}{RGB}{222,217,201}

\usepackage{bm,xstring,xfrac}   %
\usepackage{hyperref}
\usepackage{amsmath,amssymb,amsfonts}
\usepackage{color,subfigure}

\newcommand{\Q}{\mathbf{Q}}
\newcommand{\D}{\mathbf{D}}
\newcommand{\vel}{\mathbf{v}}

\def\be{\begin{equation}}
\def\ee{\end{equation}}
\def\bea{\begin{eqnarray}}
\def\eea{\end{eqnarray}}
\def\besub{\begin{subequations}}
\def\eesub{\end{subequations}}

\newcommand{\mcm}[1]{{\color{black} #1}}

\def\noi{\noindent}

\def\Q{{\bf Q}}

\def\v{{\bf v}}
\def\x{{\bf x}}

\def\nn{\nonumber}
\def\noi{\noindent}
\def\l{\left(}
\def\r{\right)}

\def\th{\theta}

\def\f{\frac}

\def\bwd{\begin{widetext}}
\def\ewd{\end{widetext}}

\def\Q{\mathbf{Q}}
\def\H{\mathbf{H}}
\def\o{\bm\omega}

\def\a{\alpha}

\def\p{\partial}

\begin{document}

\pagestyle{fancy}
\thispagestyle{plain}
\fancypagestyle{plain}{

\fancyhead[C]{\includegraphics[width=18.5cm]{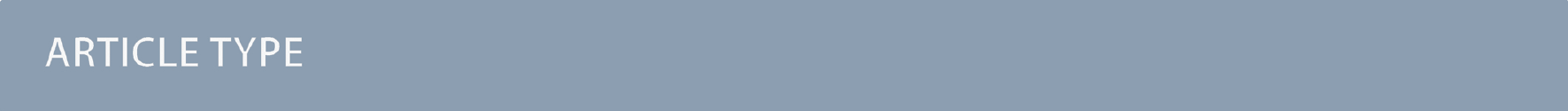}}
\fancyhead[L]{\hspace{0cm}\vspace{1.5cm}\includegraphics[height=30pt]{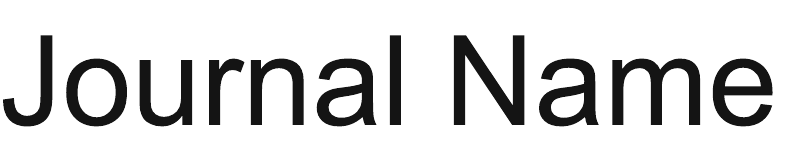}}
\fancyhead[R]{\hspace{0cm}\vspace{1.7cm}\includegraphics[height=55pt]{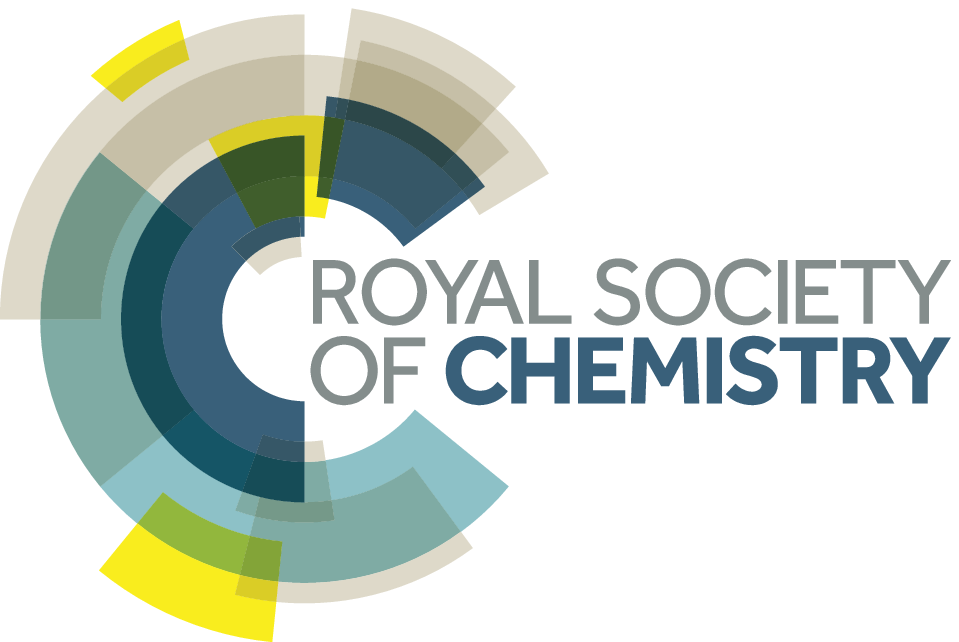}}
\renewcommand{\headrulewidth}{0pt}
}

\makeFNbottom
\makeatletter
\renewcommand\LARGE{\@setfontsize\LARGE{15pt}{17}}
\renewcommand\Large{\@setfontsize\Large{12pt}{14}}
\renewcommand\large{\@setfontsize\large{10pt}{12}}
\renewcommand\footnotesize{\@setfontsize\footnotesize{7pt}{10}}
\makeatother

\renewcommand{\thefootnote}{\fnsymbol{footnote}}
\renewcommand\footnoterule{\vspace*{1pt}%
\color{cream}\hrule width 3.5in height 0.4pt \color{black}\vspace*{5pt}}
\setcounter{secnumdepth}{5}

\makeatletter
\renewcommand\@biblabel[1]{#1}
\renewcommand\@makefntext[1]%
{\noindent\makebox[0pt][r]{\@thefnmark\,}#1}
\makeatother
\renewcommand{\figurename}{\small{Fig.}~}
\sectionfont{\sffamily\Large}
\subsectionfont{\normalsize}
\subsubsectionfont{\bf}
\setstretch{1.125} 
\setlength{\skip\footins}{0.8cm}
\setlength{\footnotesep}{0.25cm}
\setlength{\jot}{10pt}
\titlespacing*{\section}{0pt}{4pt}{4pt}
\titlespacing*{\subsection}{0pt}{15pt}{1pt}

\fancyfoot{}
\fancyfoot[LO,RE]{\vspace{-7.1pt}\includegraphics[height=9pt]{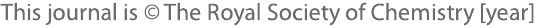}}
\fancyfoot[CO]{\vspace{-7.1pt}\hspace{13.2cm}\includegraphics{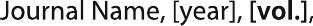}}
\fancyfoot[CE]{\vspace{-7.2pt}\hspace{-14.2cm}\includegraphics{RF}}
\fancyfoot[RO]{\footnotesize{\sffamily{1--\pageref{LastPage} ~\textbar  \hspace{2pt}\thepage}}}
\fancyfoot[LE]{\footnotesize{\sffamily{\thepage~\textbar\hspace{3.45cm} 1--\pageref{LastPage}}}}
\fancyhead{}
\renewcommand{\headrulewidth}{0pt}
\renewcommand{\footrulewidth}{0pt}
\setlength{\arrayrulewidth}{1pt}
\setlength{\columnsep}{6.5mm}
\setlength\bibsep{1pt}

\makeatletter
\newlength{\figrulesep}
\setlength{\figrulesep}{0.5\textfloatsep}

\newcommand{\topfigrule}{\vspace*{-1pt}%
\noindent{\color{cream}\rule[-\figrulesep]{\columnwidth}{1.5pt}} }

\newcommand{\botfigrule}{\vspace*{-2pt}%
\noindent{\color{cream}\rule[\figrulesep]{\columnwidth}{1.5pt}} }

\newcommand{\dblfigrule}{\vspace*{-1pt}%
\noindent{\color{cream}\rule[-\figrulesep]{\textwidth}{1.5pt}} }

\makeatother

\twocolumn[
  \begin{@twocolumnfalse}
\vspace{3cm}
\sffamily
   \begin{tabular}{m{4.5cm} p{13.5cm}}
   
    \includegraphics{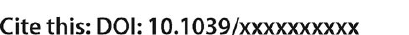} 
	& \noindent\LARGE{\textbf{Negative stiffness and modulated states in active nematics}} \\
\vspace{0.3cm} & \vspace{0.3cm} \\

    & \noindent\large{\textbf{Pragya Srivastava,\textit{$^{a\ast}$} Prashant Mishra,\textit{$^{b}$} and M. Cristina Marchetti,\textit{$^{b}$} }}\\
    
   \includegraphics{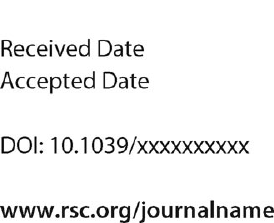} & \noindent \normalsize{We examine the dynamics of  an active nematic liquid crystal on a frictional substrate. When frictional damping  dominates over viscous dissipation, we eliminate flow in favor of active stresses to obtain a minimal dynamical model for the nematic order parameter, with elastic constants renormalized by activity. The renormalized elastic constants can become negative at large activity, leading to the selection of spatially inhomogeneous patterns via a mechanism analogous to that responsible for modulated phases arising at an equilibrium Lifshitz point. Tuning activity and the degree of nematic order in the passive system, we obtain a linear stability phase diagram that exhibits a nonequilibrium tricritical point where ordered, modulated and disordered phases meet. Numerical solution of the nonlinear equations yields a succession of spatial structures of increasing complexity with increasing activity, including kink walls and active turbulence, as observed in experiments on microtubule bundles confined at an oil-water interface.  Our work provides a minimal model for an overdamped active nematic that reproduces all the nonequilibrium structures seen in simulations of the full active nematic hydrodynamics and provides a framework for understanding some of the mechanisms for selection of the nonequilibrium patterns in the language of equilibrium critical phenomena.}\\ 

\end{tabular}

\end{@twocolumnfalse} 
\vspace{0.6cm}
 ]
\renewcommand*\rmdefault{bch}\normalfont\upshape
\rmfamily
\section*{}
\vspace{-1cm}


\footnotetext{\textit{$^{a}$~The Francis Crick Institute, Lincoln's Inn Fields Laboratory, 44 Lincoln's Inn Fields, London WC2A 3LY}}
\footnotetext{\textit{$^{b}$~Physics Department and Syracuse Soft Matter Program, Syracuse University, Syracuse, NY 13244, USA}}

\vspace{0.2in}

\noi
Active systems are continuously driven out of equilibrium by energy injected at the local scale and generate collective motion at the large scale.  Examples include bacterial colonies, in vitro extracts of cytoskeletal filaments and motor proteins, and living cells~\cite{Marchetti:2013}.
Elongated active particles can order in states with liquid crystalline order. The emergent behavior of such `living' liquid crystals has been described using liquid crystal hydrodynamics, modified to include the local energy input from active processes~\cite{Voituriez:2006,Simha:2002}. 
This work has highlighted a rich variety of collective phenomena, including spontaneous laminar flow, pattern formation, spontaneous unbinding of topological defects, and active turbulence \cite{Sanchez:2012, Giomi:2013, Giomi:2012, Thampi:2014a,Voituriez:2005, Ramaswamy:2003, Zhou:2014, Marenduzzo:2007, Thampi:2013,Giomi:2011}. 

\noi
Previous studies have established that bulk active nematics are generically unstable
 \cite{Marchetti:2013, Simha:2002, Liverpool:2006, Ahmadi:2006}. Confinement, on the other hand, has profound effects on active fluids.
It damps the flow~\cite{Liron:1976, Brotto:2013} and suppresses the generic instability of unbound systems, resulting in a finite activity threshold for the onset of spontaneously flowing states~\cite{Voituriez:2005, Ramaswamy:2007}.  Frictional damping due to confinement plays a key role in experimental realizations of  active systems.
In  bacterial suspensions  channel confinement was shown to stabilize vortex structures, leading to a state with a steady unidirectional circulation~\cite{Wioland:2013}.

\noi
\mcm{In confined nematics the energy input from active stresses is dissipated  both via viscous flows that mediate hydrodynamic interactions between the active units and via friction with a substrate. Systems where viscous dissipation dominates so that frictional damping is negligible and momentum is conserved are referred to as `wet', while those where dissipation is mainly controlled by friction are referred to as `dry'.}  
\mcm{Previous work has examined both wet ~\cite{Giomi:2013, Thampi:2013, Thampi-RS,Thampi:2014a}  and dry ~\cite{Putzig:2016, Shi:2013,Oza:2016,Thampi:2014b,Thampi:2015a} systems, as well as the crossover between the two~\cite{Amin:2016}, revealing a rich dynamics driven by orientational instabilities and  by the unbinding and proliferation of topological defects. A detailed summary of recent work most closely related to ours is given in Section \ref{compare}.  A remarkable phenomenon reported by several authors is the orientational ordering  of the axes of the comet-like  $+1/2$ disclinations. 
In experiments in suspensions of microtubule bundles the $+1/2$ defects were observed to organize in nematically ordered states~\cite{DeCamp:2015}. Numerical studies of dry systems have reported both nematic ~\cite{Oza:2016} and polar~\cite{DeCamp:2015,Putzig:2016} order of the defect orientations and a remarkable defect-ordered state accompanied by a flow-vortex lattice at the crossover between wet and dry regimes~\cite{Amin:2016}. This range of results indicates that more work is needed to understand what controls the nature of defect ordering in these systems. }

Finally, the dynamics of  dry \mcm{polar~\cite{Wensink:2012} and nematic~\cite{Oza:2016}} active fluids has also been studied via minimal models where  pattern formation is driven by diffusive currents in the order parameter equations, \mcm{with the assumption of a negative order parameter stiffness, which provides an appealing and simple explanation for the instability mechanisms. Recently, it has been shown that in systems with polar symmetry such a negative stiffness can arise from competing aligning and antialigning interactions~\cite{Grossman:2014} or from the interplay of polar alignment  and intermediate range hydrodynamic interactions~\cite{Heidenreich:2015}. }

\noi
In the present work, we consider a dry active nematic fluid, where the energy input from active stresses is balanced solely by frictional damping. By eliminating the flow velocity in favor of active stresses we obtain a single equation for the nematic order parameter, with elastic constant renormalized by activity. We show that activity can drive such elastic constants to negative values, providing a mechanism for pattern formation 
capable of describing in a unified manner all the spatio-temporal structures obtained in previous models. 
the derivation of a single equation for the nematic order parameter tensor, with elastic constants  renormalized by activity. 
A linear stability analysis of  the homogeneous 
isotropic and ordered states highlights the analogy with the onset of modulated states at a Lifshitz point in equilibrium systems and identifies the  length scales that control stable patterns. We construct a linear stability phase diagram (Fig.~\ref{Linear_plots}) and show that all linear stability boundaries meet at a single point that plays the role of a nonequilibrium multicritical point. 
The linear analysis is supported by numerical integration of the nonlinear equations that yield a succession of spatially and temporally inhomogeneous structures of growing complexity with increasing activity.

\noi
In Section \ref{Equations} we introduce the hydrodynamic model of a dry active nematic and show that it can be reduced to  a single equation for the nematic order parameter tensor by eliminating the flow. We then discuss the linear stability phase diagram in Section \ref{linear},  and  present the
results of the numerical integration of the non-linear equations in Section \ref{numerics}. We conclude with a comparison with previous work and a brief discussion.

\section{Active Nematics on a Substrate}
\label{Equations}
\noi
We consider a thin layer of active nematic liquid crystal on a substrate providing frictional damping. 
This geometry is motivated by experiments on suspensions of microtubule bundles at an oil water interface~\cite{Sanchez:2012,DeCamp:2015}, although it does not incorporate the full hydrodynamics of the system as described for instance in Ref.~\cite{Gao:2015}. \mcm{We focus on high density of active units and consider an effective one-component fluid description, as done in much of the previous literature \cite{Kruse:2005, Fielding:2011}. We assume a constant density, but do not enforce incompressibility, i.e., $\bm \nabla
\cdot \v \not=0$.} 
\mcm{This can be realized in systems where density conservation is broken, for instance, by birth/death processes or because active units can enter or leave the nematic layer through the surrounding bulk fluid. }
The long-wavelength dynamics  is  then formulated  in terms of the nematic tensor order parameter $\Q$, of components $Q_{ij}=S\l n_i n_j - \f{\delta_{ij}}{2}\r$ with $\mathbf{n}$ the director field, and the flow velocity, $\mathbf{v}$. \mcm{The absence of the incompressibility constraint allows us to directly eliminate the flow to obtain a minimal model in terms of $\mathbf{Q}$ dynamics only. 
We have also verified that imposing incompressibility only changes the location of the linear stability boundaries between the various regimes, but leaves the results qualitatively unchanged (see Appendix B). }

\noi
The  dynamics is then governed by an equation for the alignment tensor, coupled to the Stokes equation describing local force balance,
\bea
&&\left(\partial_t+\v\cdot\bm\nabla\right)\Q = \mathbf{S}(\mathbf{Q},\mathbf{A}) + \frac{1}{\gamma} \H\;,
\label{Qeqn1}\\
&&\Gamma\mathbf{v}=\eta\nabla^2\v+\alpha\bm\nabla\cdot\mathbf{Q}\;.
\label{eq:v}
\eea
The first term on the right hand side of Eq.~\eqref{Qeqn1} is a symmetric traceless tensor that describes the coupling of orientation and flow.   In two dimensions it is given by 
\be
\label{eq:S}
\mathbf{S}= \lambda\left (\mathbf{D} -\frac12{\rm Tr}\mathbf{D}\right)+ \Q\cdot\bm\o-\bm\o\cdot \Q-\lambda\mathbf{Q}{\rm Tr}[\mathbf{Q}\cdot\mathbf{D}]\;,
\ee
%
where $\mathbf{D}$  and $\bm\omega$ are the symmetric and antisymmetric strain rate tensors, $D_{ij}=(A_{ij}+A_{ji})/2$ and $\omega_{ij}=(A_{ij}-A_{ji})/2$, with $A_{ij}=\partial_iv_j$ and $\lambda $ the nematic flow alignment parameter. \mcm{It should be noted that the presence of a substrate breaks Galilean invariance. As a result, the coefficients of the convective term on the left hand side of Eq.~\eqref{Qeqn1} and of the coupling of orientation to vorticity in Eq.~\eqref{eq:S} are not required to be unity. In addition, activity will modify the value of all parameters coupling orientation and flow. The general form of the $\Q$ equation is given in Appendix A. Here to decrease the number of parameters we have taken the coefficients of the advective and vorticity coupling to be $1$ and have introduced a single flow coupling parameter $\lambda$ as appropriate for passive $2d$ nematics.
%
%
Note that with this choice $\lambda$ is known to be finite in the limit where the strength $S$ of nematic order vanishes~\cite{Forster:1974,Kuzuu:1984,Stark:2003}. }

\noi
The second term on the right hand side of Eq.~\eqref{Qeqn1} is the molecular field, $\H=-\left(\delta F/\delta \Q\right)^T$, where the superscript $T$ denotes the traceless part of the tensor, $\gamma$ is the rotational friction, and $F$ is the  Landau-de Gennes free energy of a two-dimensional nematic \cite{DeGennes:1993} , given by
\bea
F = \frac12\int_{\bf r} \Big\{ A(1-r) {\rm Tr}\Q^2+A r \left({\rm Tr}\Q^2\right)^2 + K \l \p_iQ_{jk}\r^2
+\kappa( \partial_i\partial_jQ_{kl}) ^2\Big\}\;.
\label{F2d}
\eea

\begin{figure}[h]
\hspace{-2cm}
{\includegraphics[scale=0.3]{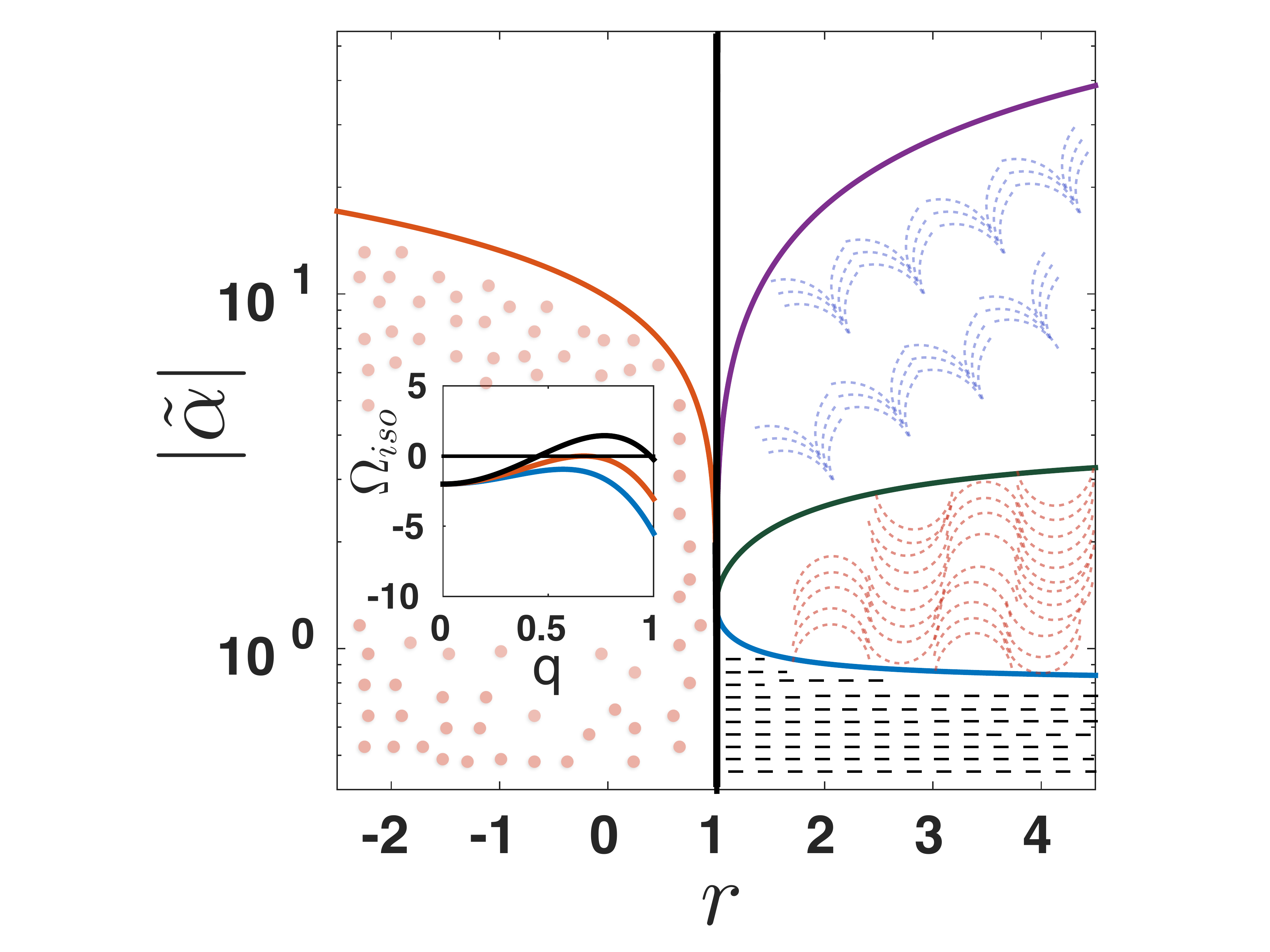}}
\caption{Phase diagram in the $(r,|\tilde\alpha|)$ plane obtained by linear stability analysis.  The vertical black line at $r=1$ is the mean-field transition between the disordered  and ordered  nematic states. The disordered state ($r<1$) becomes unstable via an isotropic instability at the value of activity given in Eq.~\eqref{a_iso} (orange line). The ordered state ($r>1$)  first becomes unstable to the growth of bend fluctuations of the director field (blue line,  Eq.~\eqref{a_th} for $\psi=0$). We  also show the lines corresponding to the change in sign of the growth rate of splay fluctuations of the director (green line, Eq.~\eqref{a_th} for $\psi=\pi/2$) and the magnitude of the order parameter (purple line,  Eq.~\eqref{a_S}). These additional instability lines correlate with the onset  of more complex spatial structures with increasing activity, as suggested by the shading of the various regions, and eventually the onset of active turbulence (no shading). All the instability boundaries meet at a multicritical point  where the uniform 
isotropic, uniform nematic and modulated phases coexist. Inset: the dispersion relation $\Omega_{I}$ (Eq.~\eqref{Om_iso}) that controls the growth of order parameter fluctuations for $r<1$ as a function of wavevector for three values of $|\tilde\alpha|=10, 13.26, 17$ (from bottom to top). At the transition only one critical mode $q=q_c^{I}$ is unstable. Above the transition the unstable modes lie in a band $q_-^{I}<q<q_+^{I}$. 
}
\label{Linear_plots}
\end{figure}
\noi
 The first two terms in Eq.~\eqref{F2d}, with $A>0$ an energy density scale,  control the isotropic-nematic transition in the passive system, which is continuous in mean-field in two dimensions.  
The transition occurs at  $r=1$. For $r<1$ the ground state of the free energy is a disordered state with $S=0$. For  $r>1$ the ground state is an ordered nematic state with $S=S_0=\sqrt{(r-1)/r}$. 
 The third term in $F$  is the elastic energy cost for deformation of the order parameter, with $K$ a Frank constant. For simplicity we use a single elastic constant approximation which equates the cost of splay and bend deformations~\cite{DeGennes:1993,BerisEdwards:1994}.  The last term in Eq. \eqref{F2d} 
describes an effective surface tension of phenomenological strength $\kappa$, assumed isotropic for simplicity. This term provides stability at small length scales.  The molecular field is then given by
\bea
\mathbf{H}&=&- A\left[ 1-r + 2 r {\rm Tr~}\Q^2\right] \mathbf{Q} + K \nabla^2 \mathbf{Q} -\kappa \nabla^4 \mathbf{Q}\;.
\label{Hexp}
\eea
In the Stokes equation, Eq.~\eqref{eq:v}, $\Gamma$ is the friction from the substrate, $\eta$ the shear  viscosity, and $\alpha$ is a measure of activity, with dimensions of stress: $\alpha>0$ corresponds to contractile active stresses, as obtained for instance in monolayers of fibroblasts~\cite{Duclos:2014},  while $\alpha<0$ corresponds to extensile stresses, as obtained in  suspensions of microtubule bundles~\cite{Sanchez:2012} or \emph{Bacillus subtilis} swimming in a nematic liquid crystal~\cite{Zhou:2014}. We have neglected   elastic stresses that yield terms of  higher order in the gradients of the alignment tensor.

\noi
We focus on the situation where frictional damping exceeds viscous forces, corresponding to length  scales large compared to the  hydrodynamic screening length $\ell_\eta=\sqrt{\eta/\Gamma}$. The  Stokes equation can then be written as
\be
{\bf v}\simeq \f{\alpha}{\Gamma}\bm\nabla\cdot{\mathbf Q}\;.
\label{fric_vel}
\ee
 Using Eq.~\eqref{fric_vel}, the flow velocity $\v$ can be eliminated from Eq.~\eqref{Qeqn1}
 to yield a description solely in terms of the dynamics  of the order parameter field $\mathbf{Q}$,
%
\begin{equation}
\begin{split}
\p_t \Q +\f{\alpha}{\Gamma} [(\nabla\cdot\Q)\cdot\nabla] \Q=
  -\f{A}{\gamma}\left(1-r +rS^2\right) \mathbf{Q}+\f{K}{\gamma}\nabla^2 \mathbf{Q}\\
 +\f{\alpha \lambda}{2\Gamma}  \left[ \bm \nabla \l \bm \nabla \cdot \Q\r \right]^{ST}
+\f{\alpha}{2\Gamma }\bm\Omega[\mathbf{Q}] -\f{\lambda \a}{\Gamma} B(\Q) \Q
  -\f{\kappa}{\gamma} \nabla^4 \mathbf{Q}\;,
\end{split}
\label{Qeqn}
\end{equation}
%
%
where the superscript $ST$ denotes the symmetric traceless part of a tensor, and we have used ${\rm Tr}~\Q^2=S^2/2$. The antisymmetric tensor $\bm\Omega[\mathbf{Q}]$ arises from the coupling of alignment and vorticity and has components 
$\Omega_{ij}=Q_{ik} \l \nabla_k \nabla_l Q_{lj}-\nabla_j \nabla_l Q_{lk}\r\nn- \l \nabla_i \nabla_l Q_{lk}-\nabla_k \nabla_l Q_{li}\r Q_{kj}$ and {$ B(\Q)=Q_{ij} \nabla^2 Q_{ij}/2$.
In spite of the apparent complexity of Eq.~\eqref{Qeqn},  the role of activity is transparent. The first term on the the second line of Eq.~\eqref{Qeqn} is proportional to $\nabla\nabla \Q$. Hence to linear order in $\Q$, activity renormalizes the Frank constant $K$, rendering it anisotropic and, as we will see below, driving it to zero and even negative in a range of parameters. This results in the instability of homogeneous states and the onset of spatially modulated structures via a mechanism analogue to that of an  equilibrium Lifshitz point \cite{Michelson:1977,Chaikin:1995}.
A negative value of the order parameter stiffness was assumed in Refs.~\cite{Wensink:2012,Dunkel:2013b,Oza:2016} as the mechanism responsible for driving pattern formation  in continuum models of dry polar and nematic active fluids inspired by the Toner-Tu equations. \mcm{Mechanisms that can lead to a negative stiffness have recently been identified for polar fluids~\cite{Grossman:2014,Heidenreich:2015}.  In particular, Ref~\cite{Heidenreich:2015} shows that hydrodynamic interactions mediated by active processes can drive a change in sign of the stiffness of a polar active fluid. Our work shows that in active nematic even screened flows in the overdamped limit can result in a negative nematic stiffness. } Equation~\eqref{Qeqn} has the structure of a generalized Swift-Hohenberg type equation for a tensorial order parameter, although it also includes nonlinearities usually neglected in the simplest models of this type. 
\noi
By examining  various terms of Eq.~\eqref{Qeqn}, we identify two important length scales: the nematic correlation length  $\ell_Q=\sqrt{K/\vert A(1-r) \vert}$ that diverges at the mean-field transition and
the length $\ell_\kappa=\sqrt{\kappa/K}$ obtained by balancing order parameter stiffness and tension. In the following we examine the behavior of the system by tuning the parameter $r$ that drives the system across the mean-field isotropic-nematic transition and the activity $\alpha$. We choose $\ell_Q^0=\sqrt{K/A}$ as  unit of length and $\tau=\gamma/A$ as unit of time. Eq.~\eqref{Qeqn} can then be written in terms of dimensionless quantities as
%
\begin{equation}
\begin{split}
\p_t \mathbf{Q} +\tilde{\alpha} [(\nabla\cdot\Q)\cdot\nabla] \mathbf{Q}=-\left(1-r + rS^2\right) \mathbf{Q}+ \nabla^2 \mathbf{Q}\\ +\f{\tilde{\alpha} \lambda}{2}  \left[ \bm \nabla \l \bm \nabla \cdot \Q\r \right]^{ST}+\f{\tilde{\alpha}}{2}\bm\Omega[\mathbf{Q}] -\lambda \tilde{\a} B(\Q) \Q - \tilde{\kappa}\nabla^4 \mathbf{Q}\;.
\label{Dim_less}
\end{split}
\end{equation}
%
\noi
\mcm{Several other groups have  studied  overdamped active nematics. 
We present a summary of previous work and comparison to our model  in Section \ref{compare}}.
Our minimal model of active nematics is controlled by only three dimensionless parameters: $r$, which controls the 
passive transition between uniform isotropic and ordered states,  a dimensionless activity $\tilde{\a}=\alpha \gamma/\Gamma K$, and  a dimensionless tension
$\tilde\kappa=\kappa A/K^2=(\ell_\kappa/\ell_Q^0)^2$.
Additionally the behavior depends on
the flow alignment parameter, $\lambda $. This  microscopic parameter is controlled by molecular shape and degree of nematic order, with  $\vert \lambda\vert >1$ corresponding to flow alignment of the nematic director in a 
shear flow and $\vert \lambda \vert <1$ corresponding to flow tumbling \cite{Forster:1974}. Below we focus on
elongated self-driven units, such as microtubule bundles or bacteria, where $\lambda$ is positive, and consider the flow aligning case, hence restrict ourselves to   $\vert\lambda\vert >1$. 

In the next section we examine the linear stability of both isotropic and nematic states and show that activity can render both unstable.

\section{Linear stability of homogeneous states and ``multicritical point''}
\label{linear}

 \noi
In this section we examine the linear stability of the homogeneous, steady state solutions of Eq.~\eqref{Qeqn} by considering the linear dynamics of fluctuations $\delta\Q=\Q-\Q^0$ of the order parameter from its uniform value, $\Q_0$. The disordered state ($r<1$) has $ \Q^0=0$.  In the ordered state ($r>1$) 
we choose the $x$ axis along the direction of broken symmetry,  so that $Q_{ij}^0= \f{S_0}{2} \l \delta_{ix} \delta_{jx}-\delta_{iy} \delta_{jy}\r$, with $S_0 = \sqrt{(r-1)/r}$. In this section we use dimensionful  quantities to better highlight the physics of the various instabilities.

\begin{figure*}
\hspace{-1cm}
{\includegraphics[scale=0.23]{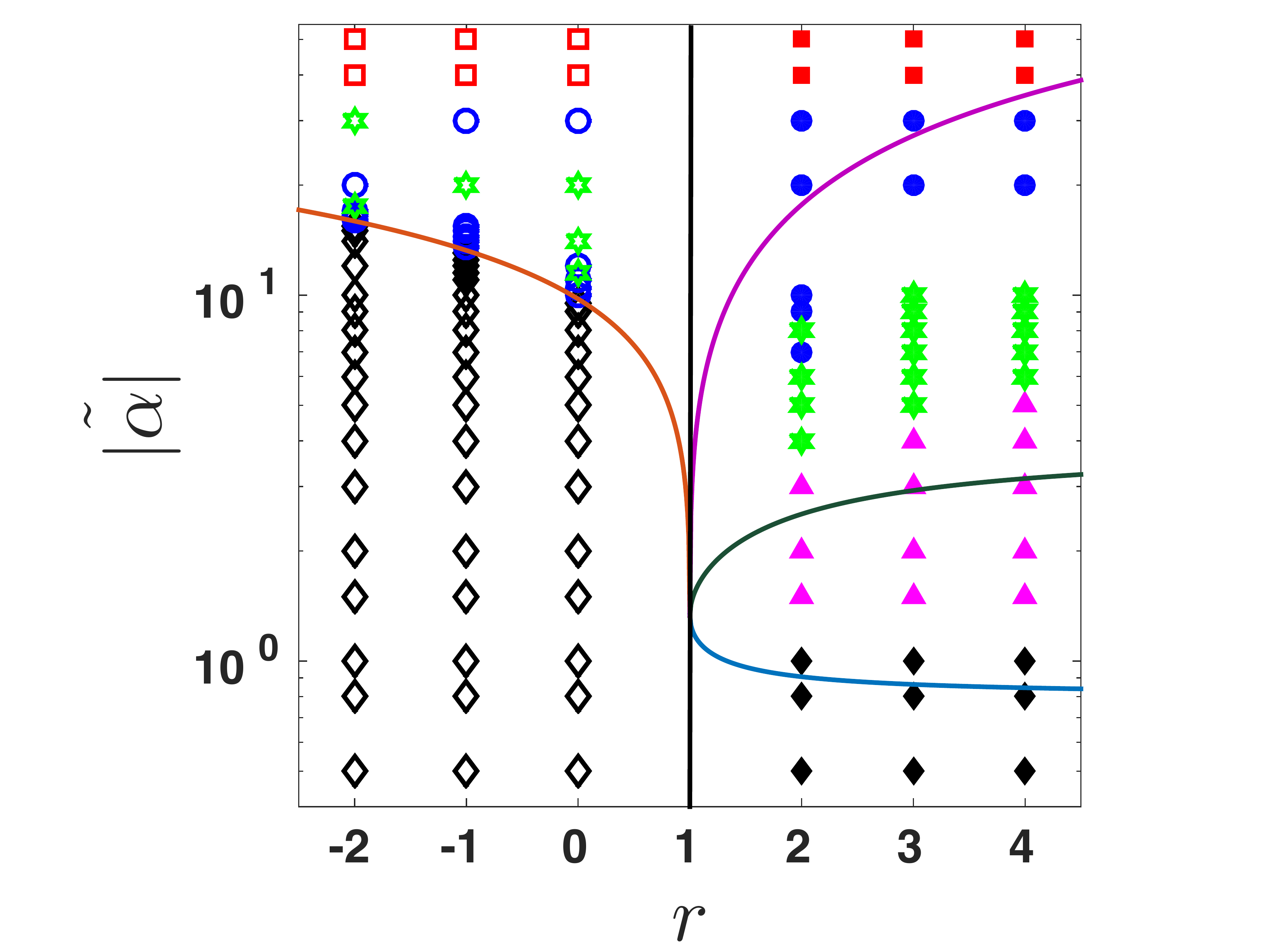}}
{\includegraphics[scale=0.27]{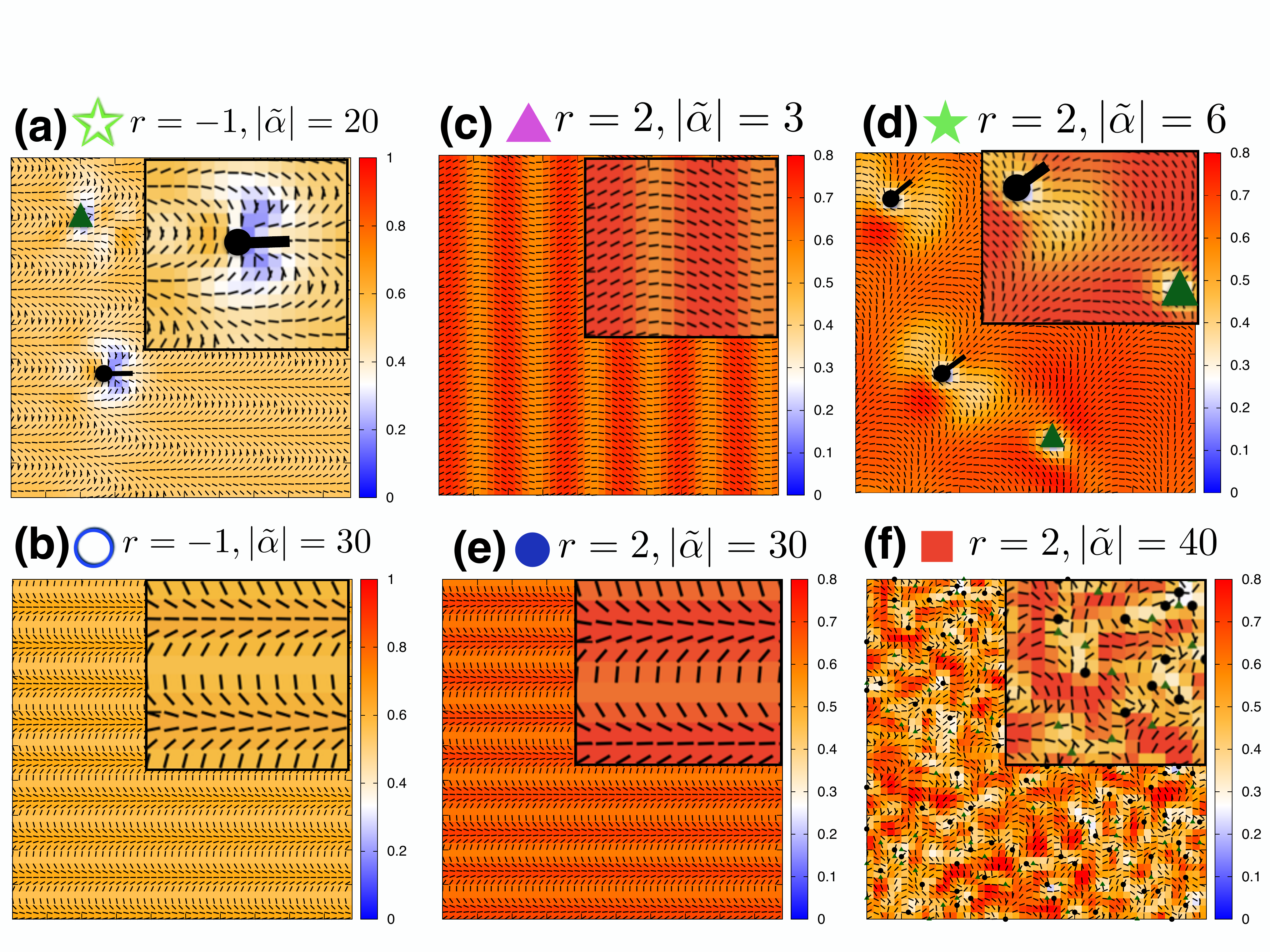}}
\caption{ Numerical phase diagram  and snapshots of the various steady states for $\tilde{\kappa}=10$ and $\lambda=1.5$.
\emph{Left panel}: Numerical phase diagram in the $(r,|\tilde\alpha|)$ plane.
The vertical black line at $r=1$ is the mean-field I-N transition from  the isotropic ($r<1$) to the nematic ($r>1$) state. 
The symbols refer to various steady state configurations obtained by numerical solution of the nonlinear equations and quantified as described in the text. Open symbols are used for states with $r<1$ and filled symbols for states with $r>1$, as follows:
uniform isotropic state (open black diamonds),  uniform nematic (filled black diamonds ), kink walls with defects  (green stars), kink walls with no defects
 (blue circles),  
bend modulations (filled magenta triangles),  turbulent state (red squares). The solid  lines denote the various  linear stability boundaries described in Fig.~\ref{Linear_plots}.
Figures (a-f) show typical snapshots of the various states:  kinks walls with (a and movie S1 in SI) and without (b and movie S2 in SI) defects  for 
 $r<1$; (c and movie S3 in SI) bend modulations, (d and movie S4 in SI) kink walls  with defects, 
(e and movie S5 in SI) kink walls  with no defects, and (f and movie S6 in SI) turbulent state.  Snapshots (c-f) are for $r>1$.  The color scale describes the magnitude of the order parameter $S$ (the mean field value is  $S_0=0.78$ in (c) to (f)). The black lines denote the direction of the director. Each snapshot also shows a blow up of the structure in the upper right corner. Disclination defects, when present, are denoted by green filled triangles ($-1/2$) and  black filled circles ($+1/2$),  with a short black solid line indicating the tail of the comet-like $+1/2$ defect. In the extensile case considered here the $+1/2$ defects move in the direction of the black dot.  } 
\label{Phdia_ststates}
\end{figure*}

\subsection{Isotropic state - $r<1$}
\noi
In this case $\Q^0=0$ and the linear dynamics of the fluctuations is controlled by the equation
\bea
\gamma\p_t \delta {\Q} = -A(1-r) \delta {\Q} + K_{I}(\alpha)\nabla^2 \delta {\Q} -\kappa\nabla^4 \delta {\Q}\;,
\label{per_iso}
\eea
where 
\bea
\label{Kiso}
K_{I}(\alpha)=K+\frac{\alpha \lambda\gamma}{2 \Gamma}
\label{K_iso}
\eea 
is an effective elastic constant renormalized by activity. By expanding the fluctuation in Fourier components as $\delta {\bf Q} = \int_{\bf r} {\bf Q}_{\bf q}(t) \exp(i {\bf q} \cdot {\bf r})$, we immediately obtain the isotropic dispersion relation of the mode that controls the growth rate of fluctuations as 
%
\bea
\Omega_{I} = - \left[A(1-r) + K_{I}(\alpha)q^2 + \kappa q^4\right]/\gamma\;.
\label{Om_iso}
\eea
The evolution of the dispersion relation  with activity is shown in the inset of Fig.~\ref{Linear_plots}. The  effective stiffness $K_{I}$ can become negative at large activity  provided $\alpha \lambda<0$, corresponding to either axially elongated active units ($\lambda>0$) with extensile activity ($\alpha<0$) or oblate active units ($\lambda<0$) with contractile activity ($\alpha>0$). The duality between elongated pushers and oblate pullers has been discussed before in `wet' active fluids \cite{Giomi:2010}. Here we will restrict ourselves to the case of elongated extensile swimmers with $\alpha\lambda<0$. In this case the relaxation rate $\Omega_{I}$ becomes positive for $K_{I}<-K_c$, with $K_c=2\sqrt{A\vert1-r\vert\kappa}$, corresponding to
\bea
\vert {\a}\vert > \alpha_{cI}=\frac{2\Gamma}{\gamma \lambda}\l K+K_c \r \;,
\label{a_iso}
\eea 
and for a band of wave vectors $q_{-}^{I}\leq q\leq q_{+}^{I}$, with 
\bea
q_\pm^{I}=\left[\f{1}{2\kappa}\l-K_{I}(\a)\pm \sqrt{[K_{I}(\a)]^2-K_c^2}\r\right]^{1/2}\;.
\eea
 Note that $K_c$ vanishes at the mean-field IN transition ($r=1$).
 The wave vector of the fastest growing mode, corresponding to the maximum of the dispersion relation shown in the inset of Fig.~\ref{Linear_plots}, is $q_m^I=\sqrt{\vert K_{I}(\alpha)\vert/(2\kappa)}=\sqrt{K_c+\gamma\lambda(\vert\alpha\vert-\alpha_c)/(2\Gamma)} \sim (\vert\alpha\vert-\alpha_c)^{1/2}$, where the last approximate equality holds near the mean-field IN transition, where $K_c=0$.   The fastest growing mode controls the length scale of the spatially modulated states.  At onset ($\vert\alpha\vert=
\alpha_c$) a single mode of wave vector $q_c^I=q_m^I(\a_c)=\sqrt{K_c /(2\kappa)}$ becomes unstable. Away from the onset of the instability, where $|K_{I}|>K_c$, we find $q_-^{I}\simeq\sqrt{A/|K_{I}|}$ and $q_+^{I}\simeq\sqrt{|K_{I}|/\kappa}$.   The instability is driven by activity that softens the elastic energy cost of order parameter fluctuations, allowing the formation of ordered regions in the isotropic state. 
At long wavelengths ($q<q_-^I$) this destabilizing effect is balanced by the passive damping of the order parameter} controlled by the rate $A/\gamma$. At short wavelengths ($q>q_+^I$) it is balanced by the surface tension $\kappa$ associated with the creation of isotropic/nematic interfaces. The threshold for the instability shifts to higher magnitudes of activity as one moves deeper into the disordered state. This mechanism for pattern formation is formally analogue to that described in Ref.~\cite{Cates:2010} for reproducing micro-organisms with a density-dependent diffusivity that can be driven to zero or even to negative values  by the suppression of motility due to crowding. The instability of an isotropic suspension of pullers has been discussed before for polar swimmers \cite{Ezhilan:2013}, but it is a new result of this work in the context of nematics.  

 \subsection{Ordered state - $r>1$}

 \noi
We now examine the stability of the homogeneous ordered state for $r>1$.
We write directly linear equations for the Fourier components of $\delta Q_{xx} $ and $\delta Q_{xy}$, given by
\bea
&&\gamma\p_t  Q_{xx}(\mathbf{q}) = -\left[2A(r-1) + K_S(\alpha) q^2 + \kappa q^4 \right] Q_{xx}(\mathbf{q})
\label{Qxlin}\;\\
&&\gamma\p_t  Q_{xy}(\mathbf{q})=-\left[ K_\theta(\alpha,\psi)q^2+ \kappa q^4\right] Q_{xy}(\mathbf{q})+\f{\a\gamma S_0}{\Gamma} q_xq_y  Q_{xx}(\mathbf{q})\;,\nn\\
\label{Qylin}
\eea
Where 
\bea
  K_S \l \a\r &=& K +\f{\a \lambda \gamma}{2 \Gamma} \l 1- \f{S_0^2}{2}\r\;,
 \label{Ks}\\
 K_\th \l \alpha,\psi\r&=& K+\f{\a \gamma}{2 \Gamma} \l \lambda + S_0 \cos 2 \psi\r \;,
 \label{Kt}
 \eea
 with $\psi$ the angle between the wave vector $\mathbf{q}$ and the direction of order. 
  The  dynamics of fluctuations in the magnitude of the order parameter ($\sim Q_{xx}$) and the direction of broken symmetry  ($\sim Q_{xy}$)  are decoupled and the dispersion relation of the corresponding eigen-modes are given by 
 \bea
 &&\Omega_S = -\left[2A(r-1) +K_S(\a) q^2 +\kappa q^4\right]/\gamma\;,
 \label{growth_S}\\
 &&\Omega_\theta =- \left[K_\th \l \a,\psi\r q^2 + \kappa q^4\right]/\gamma\;.
 \label{growth_ang}
 \eea
 Once again, activity renormalizes the bare elastic constant 
$K$. At the mean-field transition at $r=1$, $S_0=0$ and $K_S=K_\th=K_{I}$, so that all three dispersion relations $\Omega_{I}$, $\Omega_\th$ and $\Omega_S$ merge continuously. 
Both decay rates $\Omega_\th$ and $\Omega_S$ can become positive for  flow aligning ($\lambda>1$) extensile  ($\alpha <0$) nematics, signalling the instability of the homogeneous ordered state. 

The decay rate of fluctuations in the direction of nematic order controlled by $\Omega_\th$  is positive when $K_\th(\a,\psi)<0$, corresponding to
 \bea
\vert\a\vert > \a^\th_c \l \psi\r= \f{2K \Gamma}{\gamma \l \lambda+S_0 \cos 2 \psi\r}
\label{a_th}
\eea
and the fastest growing wave vector is 
 \bea
q^\th_{max}=\sqrt{\f{\vert K_\th(\a,\psi)\vert}{2 \kappa}}\;.
\eea
The most unstable mode corresponds to bend fluctuations ($\psi=0$), as expected for extensile systems, with 
$ \a^{be}_c=\a^\th_c(\a,\psi=0)= 2 K \Gamma/[\gamma(\lambda + S_0)]$.
 
The decay rate $\Omega_S$ is positive when $K_S(\a)<-\sqrt{2}K_c$, corresponding to 
%
  \bea
 \vert\a\vert >\a_c^S=\f{2 \Gamma}{\gamma\lambda}\f{2r}{r+1}\l K+{\sqrt{2}}K_c\r\;.
  \label{a_S}
  \eea
%
The instability occurs for a band of wavevectors $q_-^S<q<q_+^S$, with
\begin{equation}
q^S_\pm=\pm\left[\frac{1}{2\kappa}\l-K_S(\alpha)+\sqrt{[K_S(\alpha)]^2-2K_c^2}\r\right]^{1/2}
\end{equation}
 and the wave vector of the fastest growing mode is $q^S_{m}=\sqrt{|K_S (\a)|/\kappa}$.

The linear stability phase diagram in the  $(r,|\tilde{\a}|)$ plane is shown in Fig.~\ref{Linear_plots} for an extensile system. All linear stability  boundaries  meet at $r=1$ and $|\a|=2K\Gamma/(\gamma \lambda)$. This represents a nonequilibrium multicritical point analogue to an equilibrium Lifshitz point, where ordered, disordered and modulated phases coexist. 
The blue line corresponds to the bend instability at $\a_c^\th(\th,\psi=0)$. Also shown is the onset of instability  $\a_c^\th(\th,\psi=\pi/2)$ of pure splay fluctuations  (green line). This  `\emph{bend-to-splay}' crossover has 
 previously been identified as the mechanism of defect formation \cite{Thampi:2014b}. Finally, the purple line is  the boundary $\a_c^S$ where fluctuations in the magnitude of the order parameter become linearly unstable. The modulated phase near onset is expected to have different structures for $r$ above and below $1$ because the instability is isotropic in wave vector for $r<1$, but dominated by bend fluctuations for $r>1$. As we will see below, this is borne out by the numerical solution of the nonlinear equations. In contractile systems, in contrast, the first linear instability is controlled by splay fluctuations.\\

 \begin{figure}
   \hspace{-0.5cm}
  \includegraphics[scale=0.28]{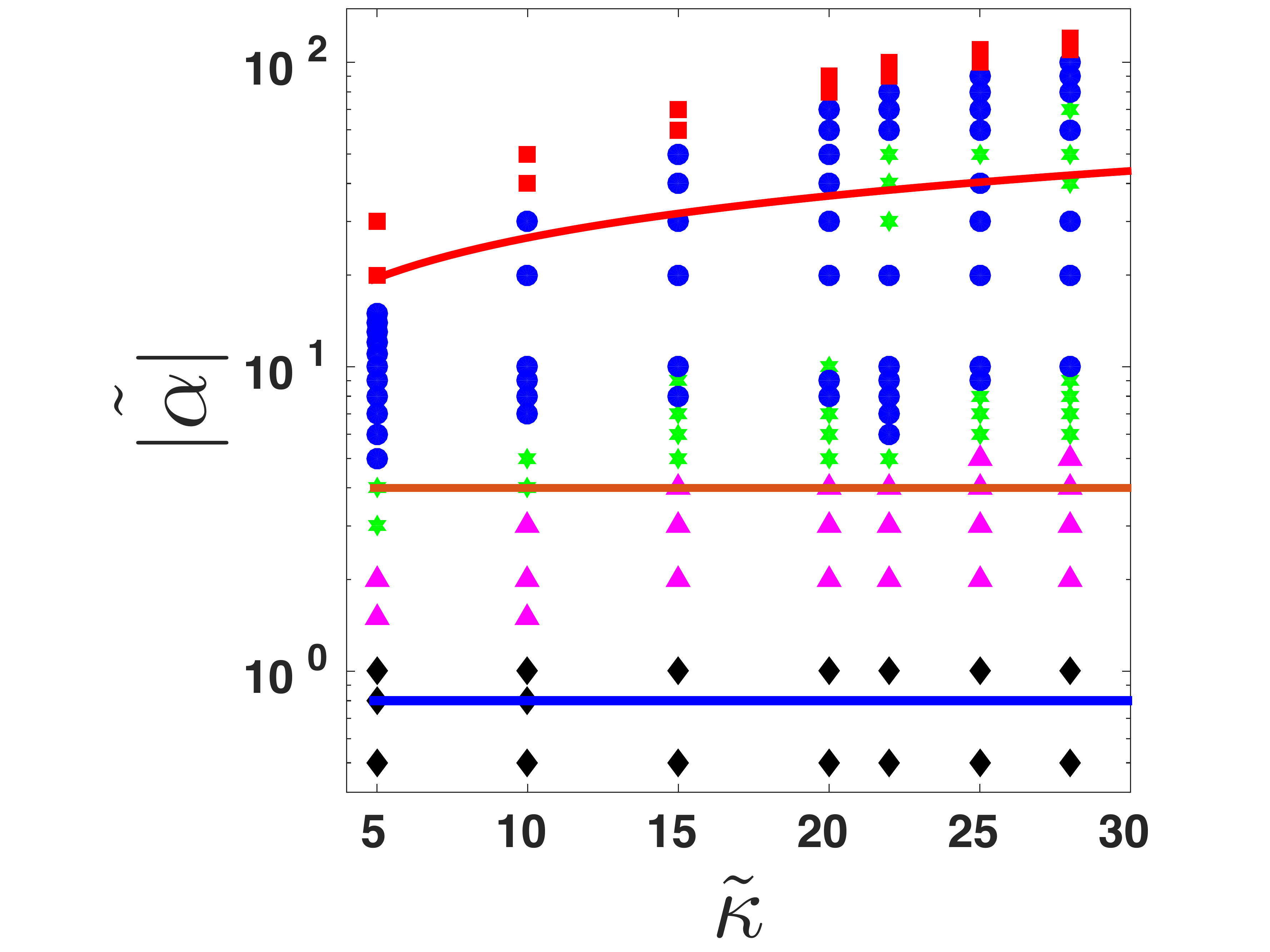}
\caption{ Numerical phase diagram in the $(\tilde{\kappa},\vert \tilde{\alpha}\vert)$ plane for $r=2$. The lines are the linear stability boundaries discussed in section \ref{linear}: bend instability (blue line), splay instability (orange line), instability of magnitude of order parameter (red line). The various symbols have same meaning as in Fig. \ref{Phdia_ststates}. 
}
\label{Ph_dia1}
\end{figure}

\section{Numerical results: kink walls, defect alignment and turbulence}
\label{numerics}

We have integrated numerically Eq.~\eqref{Dim_less} in a square box of side $L$ with bi-periodic boundary conditions. 
The $\mathbf{Q}$ dynamics is integrated numerically on a square  domain of grid size  $\Delta x= \Delta y=1$ using predictor-corrector method with the central difference scheme and a  time step $\Delta t=10^{-3}$. We
have considered  
system sizes $L=50$ and $L=100$. Most of the numerical results for $r>1$ have been obtained 
by starting the system in a uniform nematic state ordered along the $x$-axis with some superimposed  noise. We also have verified that our results are robust against the selection of initial conditions for all values of $r$.\\

\noi
We vary activity, the distance $r$ from the mean-field order-disorder transition, and the tension $\tilde{\kappa}$. As activity is increased at fixed $r$ or fixed $\tilde\kappa$, we obtain a succession of increasingly disordered structures summarized in the phase diagrams shown in Figs.~\ref{Phdia_ststates} and ~\ref{Ph_dia1}.   

To characterize the various states we have measured the local and global degree of nematic order and the number of defects.
We introduce two quantities to quantify nematic order
\bea
\langle S \rangle &=& 2\sqrt{\langle  Q_{xx}^2 + Q_{xy}^2\rangle}\;,\\
\langle Q \rangle &=& 2\sqrt{\langle Q_{xx}^2 \rangle  + \langle Q_{xy}^2 \rangle}\;,
\eea
where the brackets denote a spatial average over the system. 
Both quantities will of course be finite in a state with uniform nematic order. On the other hand, recalling that $Q_{xx}=(S/2)\cos2\th$ and $Q_{xy}=(S/2)\sin2\th$, it is easy to recognize that a state with large but randomly oriented nematic domains  will be characterized by a finite value of $\langle S\rangle$ but  a vanishing value of $\langle Q\rangle$ (in the limit of large system size).  For this reason we refer to $\langle S\rangle$ as a measure of the amount of local order in the system, while $\langle Q \rangle$ characterizes 
global nematic order at the scale of the system size.  Of course in  a uniform nematic state $\langle S\rangle=\langle Q\rangle$.

 \begin{figure}
   {\includegraphics[scale=0.28]{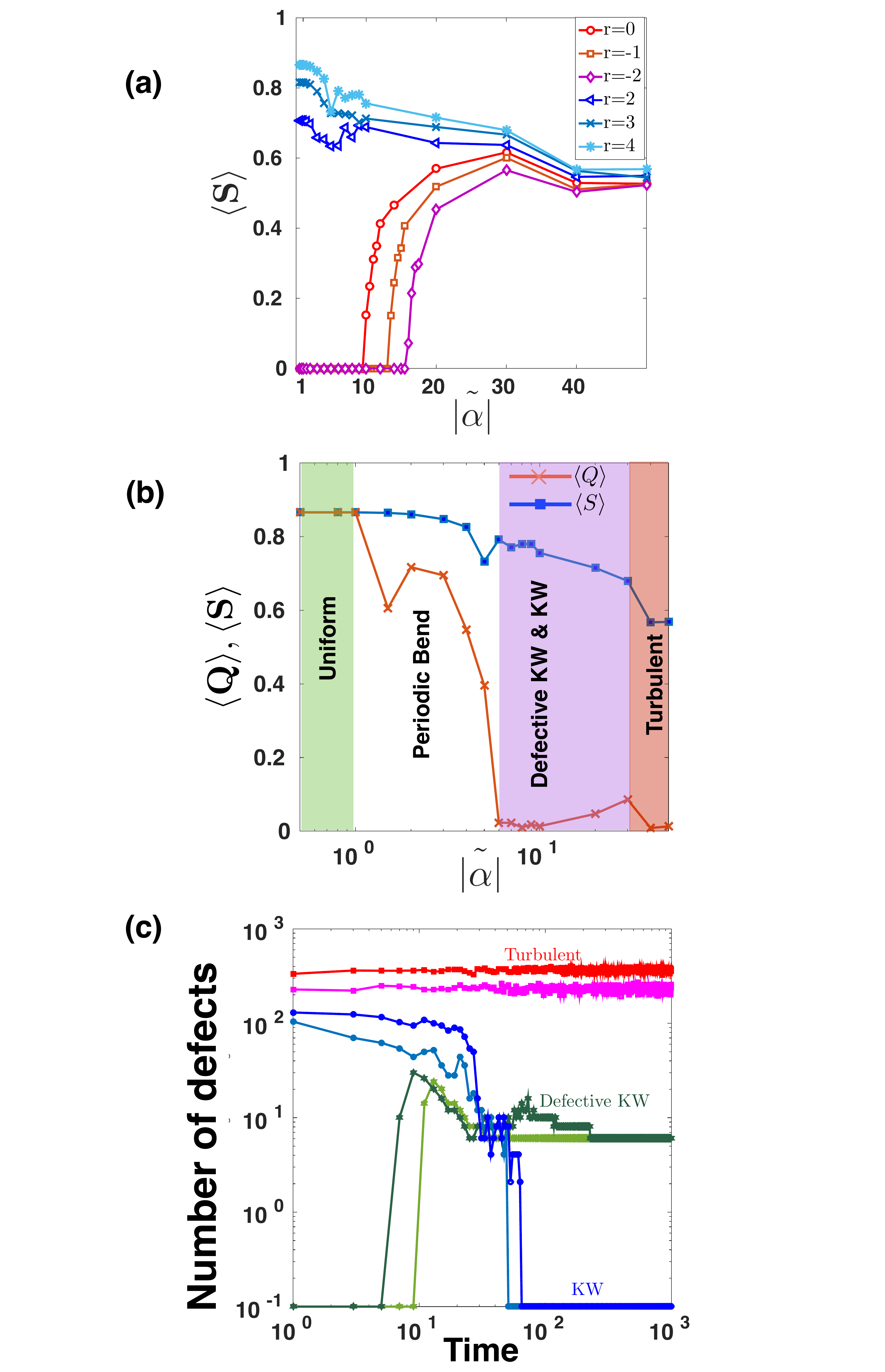}} 
   \caption{  (a) $\langle S\rangle$ as a function of activity.  The top three curves correspond to $r>1$ and show the decay of $\langle S\rangle$ with increasing activity from its mean-field value $S_0=\sqrt{(r-1)/r}$ at zero activity. The bottom three curves 
   correspond to $r<1$, where $\langle S \rangle$ ramps up sharply from its low activity value of zero at the onset $|\alpha_{cI}|$ of the linear instability. The curves through the data are a guide to the eye. 
(b) The evolution with activity of the two measure of nematic order $\langle Q \rangle$ (red crosses) and $\langle S\rangle$ (blue squares) for $r=4$, $\tilde{\kappa}=10$ and $\lambda=1.5$. The shading denotes the various regimes as estimated from the numerics. At low activity the system is in  a uniform nematic state (light green) and $\langle Q \rangle=\langle S\rangle=S_0 = 0.86$.  In the region of bend modulations (white) $\langle S \rangle$ remains close to its maximum value, while $\langle Q \rangle$ drops sharply, remaining close to zero in all subsequent regimes. The slight increase of  $\langle Q \rangle$ near the transition from KW to turbulence reflects the alignment of KW in this region. (c) Number of defects  as a function of time for six values of activity for $r=4$: 
  $|\tilde\alpha|=6,~7$, corresponding to KW with defects (green curves),  $|\tilde\alpha|=20,~30$, corresponding to defect-free KW  (blue curves),  and $|\tilde\alpha|=40,~50$, corresponding to the turbulent state (red curves). 
  }
  \label{iden_states}
\end{figure}

\begin{figure}
{\includegraphics[scale=0.18]{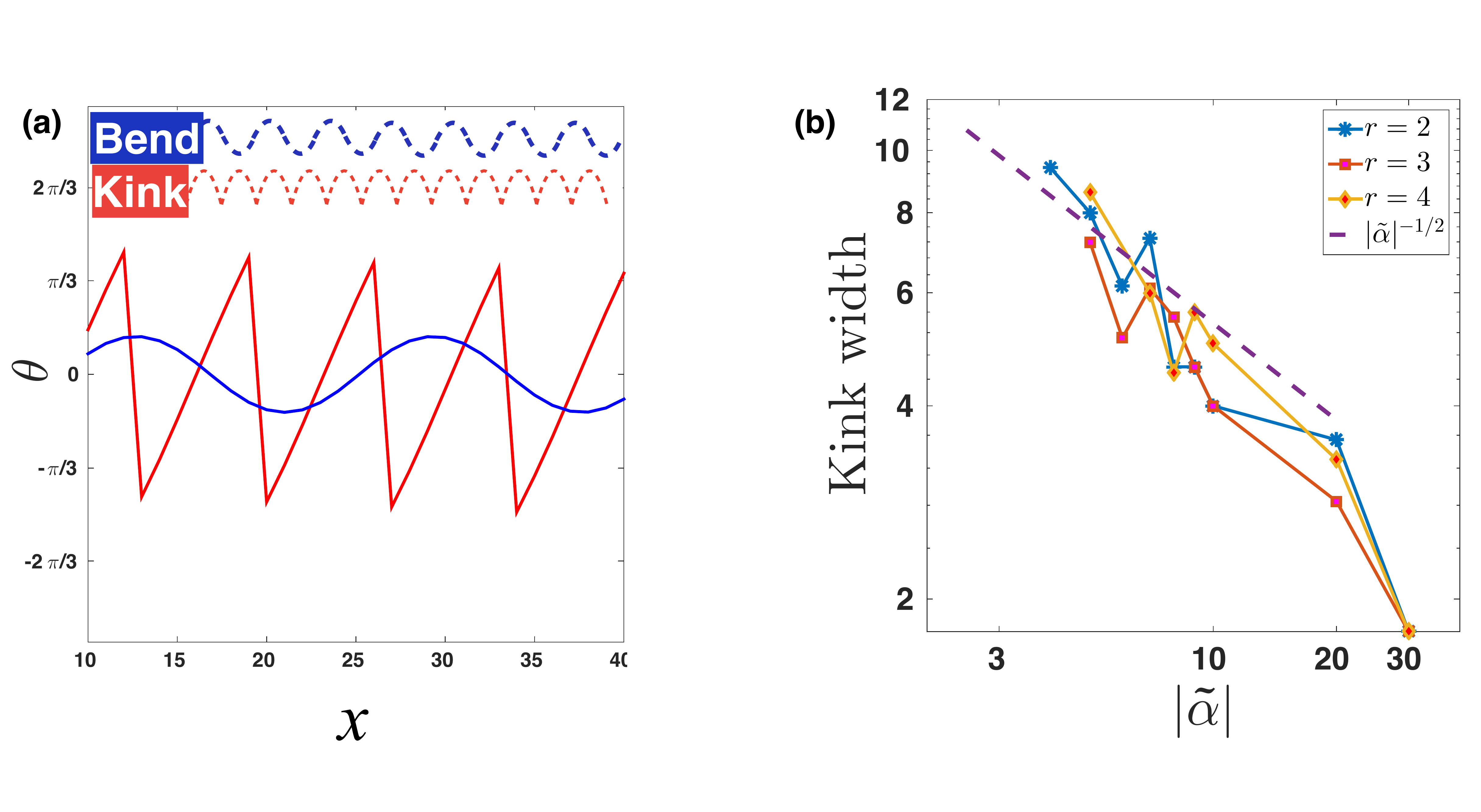}} 
\caption {(a) Profiles of the angle $\theta$ of the director with  the direction of mean order for \emph{bend} (blue curve, $|\tilde{\alpha}|=3 $) and \emph{kink walls} (red curve, $|\tilde{\alpha}|=20$), $\tilde{\kappa}=10$ and $\lambda=1.5$ The director profiles for both deformations are shown at the top of the figure. (b) Width of kink walls versus activity  for  $r=2,3,4$, $\tilde\kappa=10$ and $\lambda=1.5$.
 The dashed line has slope $-1/2$. 
 }
\label{iden_states_2}
\end{figure}

 \emph{Activity yields spatial structures}: Activity drives uniform nematic states towards more disordered configurations and builds up local nematic order in uniform isotropic states. This is displayed in  Fig. \ref{iden_states}(a) that shows the evolution with activity of local nematic order $\langle S\rangle$ for values of $r$ below  (red curves) and above (blue curves) the mean field I-N transition. At low activity the system is in a homogeneous state and $\langle S\rangle$ has its mean-field value. For $r<1$  $\langle S\rangle$ remains identically zero up to the value of activity corresponding to the linear instability of the isotropic state, and then grows with activity.  The result that activity builds nematic order may seems surprising, but is consistent with recent results for wet systems, where flows induced by active stresses were  shown to renormalize the location of the IN transition~\cite{Thampi:2015a}. \\
 
 \noi
For $r>1$ the system is in a uniform nematic state at zero activity with $\langle S \rangle=\sqrt{(r-1)/r}$. Activity disorders the system driving the onset of spatial structures of increasing complexity, and $\langle S\rangle$ decreases continuously. At large activity where the system enters the turbulent regime, the distance $r$ form the mean-field critical point ceases to provide a distinction between states and both sets of curves (  $r>1$ and $r<1$ ) converge to the same value of $\langle S\rangle$. 
 It should be stressed  that of course the line of I-N transition is itself generally renormalized  by activity via coupling to viscous flows~\cite{Thampi:2015a}. 
These effects are however, not incorporated in our model that considers 
 length scales much larger than the  hydrodynamic screening length $\ell_\eta$. 
  
We now characterize the various  regimes obtained by increasing activity.

\noi
\emph{Bend modulations:} 
When $r>1$ the uniform active nematic state becomes unstable to bend fluctuations(see movie S3 in SI) of the director at 
$\vert\a\vert>\a^{be}_c$.  This yields a state with  bands of alternating high and low value of the order parameter $S$ and associated bend deformations of the director, as  shown in Fig. \ref{Phdia_ststates}(c). The bands exhibit a high degree of smectic order. In wet systems this is the regime of spontaneous flow~\cite{Voituriez:2005}, and was reported in most previous studies with and without frictional damping~\cite{Giomi:2013, Giomi:2012,Thampi:2014a}, but not in Ref.~\cite{Oza:2016}. The activity threshold  found numerically  matches well  the value obtained from linear stability analysis, as evident from  Figs. \ref{Phdia_ststates}(a) and (b). The mean order parameter
$\langle S \rangle$  is lower than  its mean-field value but remains finite,  as shown in Fig.~\ref{iden_states}(a), while  $\langle Q \rangle$  quickly drops to $0$, as evident from
 Fig.~\ref{iden_states}(b). 
Beyond the activity where the degree of global order $\langle Q \rangle$ drops to zero,  
the system does not have a memory of the direction of initial order.  Hence, structures obtained at higher activity are expected to be the same for $r>1$ and $r<1$. This is also
  evident from the phase diagram  of Fig. \ref{Phdia_ststates}.  We discuss these states in the following.

\noi
\emph{Kink walls (KW): }
Upon increasing activity, the magnitude of $\langle S \rangle$ decreases and director deformations become larger, acquiring a splay character (see movies S2 and S5 in SI).  The system organizes in a structure of parallel kink walls, with large-scale smectic order, \mcm{resembling the patterns} seen in living liquid crystals of \textit{B. subtilis} swimming in a passive nematic~\cite{Zhou:2014,Lavrentovich:2016}. The width and spacing of the kink walls is set by the length scale $\ell_{KW}\sim (q^S_m)^{-1}\sim\sqrt{\Gamma\kappa/(|\alpha|\gamma)}\sim |\alpha|^{-1/2}$, as shown in Fig.~\ref{iden_states_2}(b) and seen in experiments~\cite{Zhou:2014}. The activity range over which the aligned KW are observed is not, however, sufficient to establish scaling. To highlight the distinction between bands of bend modulation and kink walls, we show in Fig.~\ref{iden_states_2}(a) the director profile for  the two structures.   Due to the large splay deformation of the director and the associated suppression of the local value of $S$, defect pairs unbind and travel along the kink walls (Fig.~\ref{Phdia_ststates} (c)). The majority of defect pairs annihilate at long times, as evident from the plot of number of defects versus time shown in Fig.~\ref{iden_states}(c) (dark and light green curves), resulting in a state with a small concentration of stable unbound defects (typically fewer than $10$ for $L=100$), as shown in Fig.~\ref{iden_states}(c). The few remaining $+1/2$ defects move along the kink walls  maintaining essentially parallel relative orientation (see movies S1 and S4 in SI), \mcm{i.e. exhibit polar order} as reported in~\cite{DeCamp:2015,Putzig:2016}.
  This behavior is observed in a very narrow range of activity and only for sufficiently large values of  $\tilde{\kappa}$, as shown in the phase diagram in Fig.~\ref{Ph_dia1}. 

\begin{figure*}
\centering
{\includegraphics[scale=0.23]{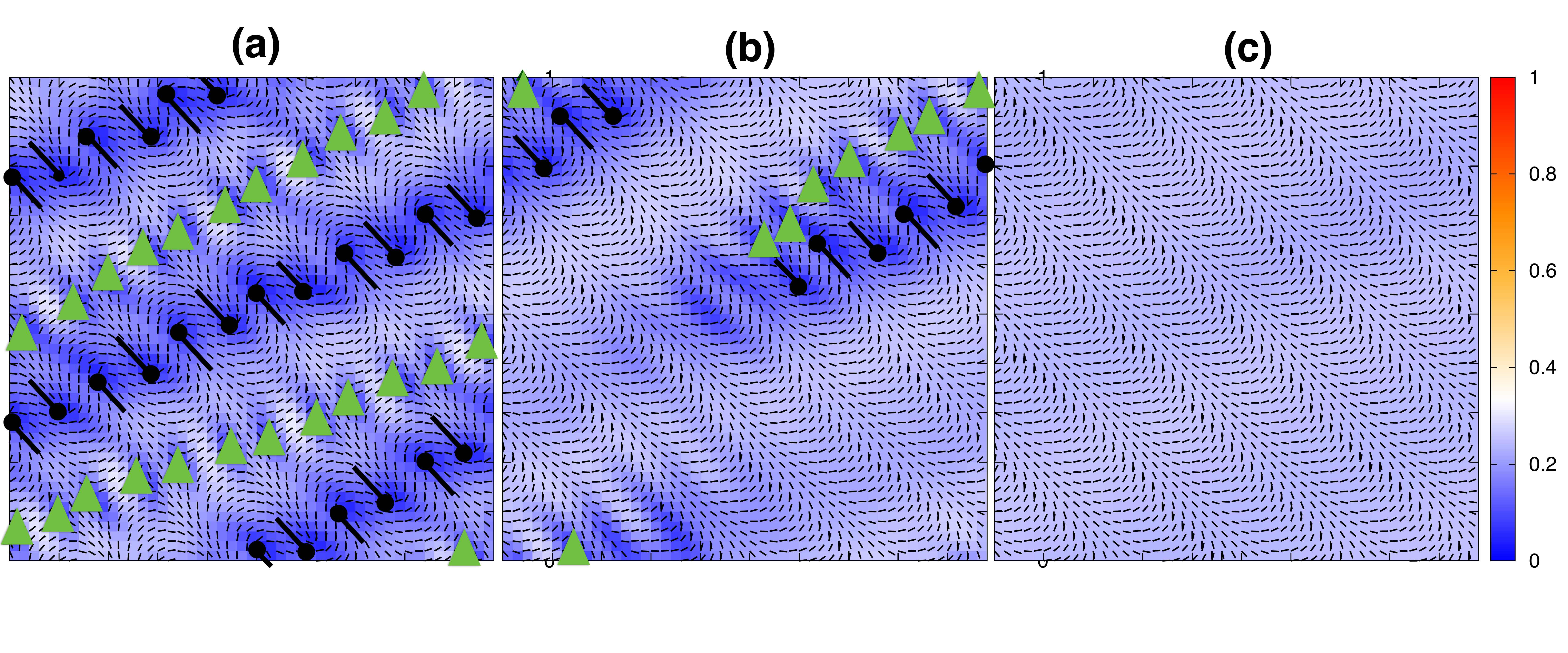}}
\caption{Snapshots at (a) t = 92$\tau$, (b) t = 143$\tau$ and (c) t = 170$\tau$ showing the transient nematic alignment of defects for $r=-1$, $|\tilde{\alpha}|=14$, $\tilde\kappa=10$ and $\lambda=1.5$(see movie S7 in SI). These structures are obtained for values of activity close to the boundary of linear stability of the uniform mean-field isotropic state, which is at $\alpha_{cI}=13.2$ for the parameters used here. The defects are highlighted using the same convention as described in the caption of  Fig.~\ref{Phdia_ststates}.
 }
\label{defect_order}
\end{figure*}

 \noi
 \emph{Defect ordering:} \mcm{As discussed in the Introduction, experiments have reported nematic order of the orientation of $+1/2$ defects in microtubule bundle suspensions~\cite{DeCamp:2015}. Previous numerical work has observed both nematic~\cite{Oza:2016} and polar~\cite{DeCamp:2015,Putzig:2016} order of the defects in dry systems and a lattice of flow vortices with rows of nematically ordered defects at the wet-to-dry crossover~\cite{Amin:2016}. In our model we have observed two types of defect-ordered structures.} 
 
\noi
For $r<1$, and in a very narrow range of activity near 
the linear stability boundary, we observe \textit{nematic} alignment of the axes of the $+1/2$ defects similar to the one seen  \mcm{ in numerical models of Refs.~\cite{Oza:2016,Amin:2016},} as shown in Fig.~\ref{defect_order}(a) and movie S7(in SI). \mcm{Unlike previous work, we find, however,} that these rows of defects are a transient state. At long times
all the defects annihilate leading to a state of kink walls, with no unbound defects, as displayed  in Fig.~\ref{defect_order} for $r=-1$ and $\vert\tilde {\a}\vert=14$. 
\mcm{Note that the work by Doostmohammadi \textit{et al.}~\cite{Amin:2016} included flow, suggesting that this may play a role in stabilizing these nematic defect rows. On the other hand, the work by Oza and Dunkel~\cite{Oza:2016} is, like ours, a dry model.  We have verified that  the rows of nematically aligned defect are long lived for the parameter values used in Ref.~\cite{Oza:2016} where many of the orientation-flow couplings are set to zero (see Appendix A). Clearly more work is needed to fully explore the very large parameter space of these rich models.}

\noi
\mcm{In addition, we observe stable \textit{polar} order of the orientation of the $+1/2$ defects in the regime of defective kink walls displayed in Fig.~\ref{Phdia_ststates}(d) and more clearly evident in Supplementary movie S4. This type of order, with defects situated at the end of kink walls, is the same as observed previously in simulations of dry systems~\cite{DeCamp:2015,Putzig:2016}.}

\noi
\emph{ Turbulent state:} At even higher activity, when the width of the kink walls become of the order of $\ell_Q^0$,
we find a turbulent state with proliferation of unbound defect (see movies S6 and S8 in SI). The evolution of the defect density 
with time is shown in Fig.~\ref{iden_states}(c) (red and orange curves).

\mcm{\section{Comparison to \mcm{previous work}}
\label{compare}

\mcm{The theoretical literature on continuum models of the emergent dynamics of active nematic liquid crystals has seen a rapid growth in recent years, with several papers directly related to the work presented here. For this reason we think it useful to summarize in this section the relation of our work  to a few others~\cite{Amin:2016,Putzig:2016,Oza:2016} that also discuss the elimination of the flow velocity  in the overdamped limit by balancing frictional  and active forces.}

\noi
\mcm{Perhaps most closely related to our work is the one by} Putzig \emph{et al.}, where the authors 
eliminate flow in favor of active stresses, but retain 
the dynamics of the density. 
In the density equation 
flow elimination leads to the 
curvature induced currents that have been studied extensively in the literature~\cite{Ramaswamy:2003}. Allowing for density variations 
yields the previously discussed phase separation in alternating bands of ordered and disordered regions 
in a very narrow region near the mean field isotropic-nematic transition [ref], but does not seem to affect the splay or 
bend instability of the order parameter which yields modulated states as found in our model.  \mcm{These authors retain general values for the various parameters that couple orientation and flow (see Appendix A for details) and  find robust polar alignment of the orientation of the $+1/2$ defects and modulated kink walls structures.
For our choice of such flow-orientation couplings, } the same stable
polar ordering of defects is obtained only in the narrow region of parameters identified by filled green stars in Fig. \ref{Linear_plots}
and Fig. \ref{Phdia_ststates}. 

\noi
\mcm{Finally, Putzig \textit{et al.} do not discuss the complete phase diagram across both the mean field isotropic and nematic states, nor highlight the  role of the negative effective stiffness as a generic pattern formation mechanism.}

\noi
A $\mathbf{Q}$-only model similar to ours was recently proposed by Oza and Dunkel~\cite{Oza:2016}, who
consider a compressible active nematic with constant density, but assume from the outset a negative value of the elastic stiffness. Additionally, these author neglect the coupling of orientation to both shear flows  ($\lambda=0$) and vorticity (see also Appendix A). For this special parameter values they find stable  nematic alignment of the $+1/2$ defects that persist in steady state at low activity. This is in contrast to our model where nematic defect ordering is always transient. \mcm{The ad hoc assumption of negative stiffness (as opposed to a stiffness that is driven negative by activity) yields a phase diagram where the homogenous states are found to be unstable even for zero value of activity.}\\

\noi
Finally,  Doosmohammadi \emph{et al}. consider an active nematic suspension with both substrate friction and viscous flow.  An intriguing result of this work is the observation of stable rows of aligned defects with nematic order 
of the $+1/2$ defects perhaps  resembling the structures reported in suspensions of microtubules \cite{DeCamp:2015}.
The ordering of defects is accompanied by an ordered lattice of flow vortices of alternating sign. 
While we observe similar states of rows of defects with nematic alignment as shown in 
Fig. \ref{defect_order}, we find that these states are transient and the steady state is defect free in this regime 
of parameters. This suggests that flow may be needed to stabilize the nematic order of defects. These authors 
also state that when frictional damping in the dominant mechanism for dissipation,
the system no longer exhibits spontaneous defect proliferation. This may follow from the fact that they 
examine behavior by increasing friction at \mcm{\emph{fixed activity},} which in the overdamped regime is equivalent 
to decreasing the effective activity $\a$ that enters in the combination 
$\a/\Gamma$.

}

\section{Conclusion and Discussion}
\label{discussion}

\noi
We have shown that at length scales larger than the 
hydrodynamic screening length a compressible active nematic film
can be described by a single equation for the tensor order parameter $\Q$.
Activity renormalizes the elastic  stiffness of the nematic and drives it to negative values, destabilizing  \emph{both} the ordered and disordered homogeneous states 
and providing a mechanism for pattern formation analogue to that at play at an equilibrium Lifshitz point. Activity plays different roles depending on the value of the parameter $r$ that tunes the proximity to the mean field isotropic-nematic transition of the passive system. For $r<1$, corresponding to a uniform passive isotropic state in mean-field,  activity builds up local nematic order, while it disrupts it  for $r>1$, which corresponds to the  ordered state in mean-field. In both cases a  moderate activity yields spatial structures  that are stabilized 
at small length scales by a surface tension that favors the formation of isotropic/nematic interfaces. 
\mcm{Our work provides a minimal model that unifies many previously presented results in a phase diagram that includes: (i) a regime of spontaneous bend deformations of the director field that corresponds to the spontaneous laminar flow state of wet systems; (ii) kink walls with unbound defects;  (iii) defect-free kink walls;  (iv) and finally chaotic dynamics with proliferation of topological defects.  The regimes of bend deformations and kink walls both exhibit large-scale smectic order, similar to the one observed in suspensions of \textit{Bacillus subtilis} swimming in chromonic liquid crystals~\cite{Zhou:2014,Lavrentovich:2016}, where substrate friction is indeed the dominant dissipation process. \\
\noi
Defect pairs unbind at the boundaries of kink walls, where nematic order distortion are largest, and zip along the walls, leaving large distortions in their trail (see SI movie S4). As activity increases, kink walls begin to meander. This eventually results in the loss of smectic order and the onset of the chaotic spatio-temporal dynamics that has been referred to as active turbulence,}
 of the value of $r$. The finding that activity builds up nematic order in the regime where the passive system is isotropic was also reported in Ref.~\cite{Thampi:2015a} for an active nematic where flow dominates over viscous dissipation. In this case it arises because flows due to active stresses effectively suppress the damping of the nematic order parameter, shifting the IN transition and enhancing the nematic correlation length. In our overdamped system the enhancement arises form nonlinear terms coupling the order parameter to active flows. 

\noi
The effective equation for the order parameter obtained here is similar to a Swift-Hohenberg equation which is known to generically give rise to 
modulated phases. \mcm{ The additional nonlinearities and anisotropies present in our model as compared to conventional Swift-Hohenberg models do not affect the symmetry of the specific nonequilibrium structures, but only the parameter values controlling the crossover between the various regimes}. In particular, the cubic nonlinearity in the term proportional to $\mathbf{B}(\Q)$ on the right hand side of Eq.~\eqref{eq:S} is necessary to ensure  that local nematic order as quantified by $\langle S\rangle$ remain bounded. Our work therefore shows that the spatially modulated structures arising in active systems, especially the kink walls,  are a generic consequence of the structure of the active hydrodynamic equations and  identifies of the softening of the elastic stiffness due to activity as a generic mechanism for pattern formation in active systems dominated by frictional damping.

\noi
Using this model, we demonstrate that activity destabilizes not only the uniform ordered state, as well established in the literature, but also the mean-field disordered state.  As well established, activity yields defect unbinding, but the proliferation of unbound defects does not grow monotonically with activity. This is evident from the sequence of steady states shown is the phase diagrams of Figs. 2 and 3: we find a state of kink walls with defects in  a range of activity that give way to defect-free kink walls at higher activity. 
Defect proliferation is the result of two competing effects: the unbinding of defects pairs that occurs mainly along kink walls and increases with activity and the pair annihilation that is largely controlled by the speed of the $+1/2$ defects, and of course defect density. Both mechanisms depend on activity and our findings suggest that one may dominate over the other in different regions of parameter space. While defect unbinding in extensile systems has been associated with a transition to regime dominated by splay deformation of the order parameter, more work is needed for a quantitative understanding of how the creation and annihilation rate depend on activity and flow.

\noi
Finally, experiments in active microtubule suspension at an oil-water interface have reported a remarkable nematic alignment of $+1/2$ defect over large scales~\cite{DeCamp:2015}. \mcm{Both nematic and polar defect alignment have been reported in previous numerical studies of various models~\cite{DeCamp:2015,Amin:2016,Oza:2016,Putzig:2016}. Here we obtain both nematic and polar alignment of the orientation of the $+1/2$ defects in different regions of parameters.  Unlike previous work, we find, however, that  nematic order is always transient when flow is completely screened by friction. Polar order is, however, stable, albeit in a narrow region of parameters.}

A more subtle interplay of viscous and frictional dissipation 
may stabilize this defect-ordered state as obtained in~\cite{Amin:2016} using a full hydrodynamic model, but more work remains to be done 
to understand the origin of this topological order.

\section{Acknowledgments}
\noi
 We thank Ananyo Maitra and Sriram Ramaswamy for insightful discussions. This work was supported by National Science Foundation through award DMR-1305184, DGE-1068780 and by the Syracuse Soft Matter Program.

\begin{appendices}
\numberwithin{equation}{section}

\mcm{\section{Overdamped nematic in $2d$ :  General equations}
\noi
Here we present the general equations for the tensor order parameter $\Q$ of an active nematic on a substrate.
The presence of the substrate breaks Galilean invariance, removing the constraints that determine the sign and value of the coefficients of the co-rotational derivative that couple alignment and flow.
In general,  the equation for the $\Q$-tensor takes the form 
\begin{equation}
\partial_t\Q+\lambda_C(\vel\cdot\bf{\nabla})\Q-\lambda_R[\Q,\mathbf{\omega}]=\mathbf{S}(\Q,\bm\nabla\vel)+\frac{1}{\gamma}\mathbf{H}\;,
\label{a1:Q}
\end{equation}
where
\bea
\mathbf{S}&=&\lambda_1\left[\D-\frac{1}{d}{\rm Tr}(\D)\right]
+\lambda'_2\left[\Q\cdot\D+\D\cdot\Q-\frac{2}{d}\mathbf{1}~{\rm Tr}(\Q\cdot\D)\right]\nn\\
&&+\lambda_2\frac{2}{d}\Q ~{\rm Tr}(\D)
-\lambda_3\Q~{\rm  Tr}(\Q\cdot\D)\;,
\label{generalS}
\eea
with $d$ the dimensionality, 
$\D$ and $\bm\omega$ the symmetrized strain rate and vorticity defined in the main text after Eq.~\eqref{eq:S} and $[\mathbf{A},\mathbf{B}]=\mathbf{A}\cdot \mathbf{B}-\mathbf{B}\cdot \mathbf{A}$. In system with Galilean invariance 
the coefficients of the advective derivative and the orientation-vorticity coupling are constrained to have unit value, or
\begin{equation}
\lambda_C=\lambda_R=1\;.
\end{equation}
Additionally, in passive liquid crystals the other parameters coupling orientation and flow  have been calculated ~\cite{Stark:2003} and are generally taken as
\begin{eqnarray}
&&\lambda'_2=-\lambda_2=1\;,\notag\\
&&\lambda_1=\lambda_3=\lambda\;.
\end{eqnarray}
In $2d$ the fact that $\Q$ is a symmetric traceless tensor yields the identity 
\begin{equation}
\Q\cdot\D+\D\cdot\Q-\mathbf{1}~{\rm Tr}(\Q\cdot\D)=\Q~{\rm Tr}(\D)\;.
\label{identity}
\end{equation}
Thus 
%
%
for a $2d$ active nematic on a substrate 
the matrix $\mathbf{S}$ can be written without loss of generality as
\begin{equation}
\mathbf{S}=\lambda_1 \left[\D-\frac{1}{d}~{\rm Tr}(\D)\right] +\lambda_2\Q~{\rm Tr}(\D)-\lambda_3 \Q ~{\rm Tr}(\Q\cdot\D)\;.
\label{ourmodel}
\end{equation}
In order to reduce the number of parameters in our work we have taken $\lambda_C=\lambda_R=1$, $\lambda_2=0$ and $\lambda_1=\lambda_3=\lambda$.  
\noi
We now make an explicit comparison between our model and the equations used in a few other recent works that are most directly relevant to ours. 

\noi
\textit{Oza and Dunkel~\cite{Oza:2016} : } Like us, these authors have considered a compressible, overdamped active nematic with 
constant density. These authors have chosen $\lambda_C=-\lambda_2=1$ and  $\lambda_1=\lambda_3=0$. The parameter $\lambda_R$ (denoted as $\kappa$ in their work) is retained as arbitrary in the equations, but set to zero in the numerical calculations. As a result, the only term coupling orientation and flow is a convective term of the form $\bm\nabla\cdot(\v\Q)=(\v\cdot\bm\nabla)\Q+\Q(\bm\nabla\cdot\v)$. The only active terms then come from eliminating the flow using Eq.~\eqref{fric_vel}. These terms cannot yield the activity-induced suppression of the nematic stiffness, which is therefore simply assumed to be negative from the outset. One issue with this formulation is that the model predicts that the homogeneous states are unstable even at zero activity.

\noi
\textit{Putzig \textit{et al.}~\cite{Putzig:2016} : } These authors consider an overdamped active nematic and eliminate flow using Eq.~\eqref{fric_vel}, but allow the density to vary, although it is a little unclear whether their density should be interpreted as the concentration of active units or the density of the suspension.  They  work in $2d$ and consider arbitrary values for the parameters coupling concentration and flow.  Taking advantage of the identity \eqref{identity}, and denoting with a subscript $P$ the parameters of Ref.~\cite{Putzig:2016} when different from ours, their equations correspond to the following choice: $\lambda_2=\lambda_2^P-\lambda_C$, $\lambda_3=0$, with $\lambda_1$, $\lambda_C$ and $\lambda_R$ allowed to have generic values.

\noi
\textit{Doostmohammadi \textit{et al.}~\cite{Amin:2016} : } These authors consider an active nematic  \textit{suspension} in $3d$ and include both viscous flows and friction with a substrate.  All the parameters coupling orientation and flow are assumed to have the value they would have in a passive system with Galilean invariance. They assume the total density of the suspension to be constant and require the flow to be incompressible, but allow the concentration of active particles to vary.

\section{LInear Modes for general flow coupling parameters}
In this Appendix we discuss the linear modes obtained by using the general form of the equations for an overdamped nematic as given in Eqs.~\eqref{a1:Q} and \eqref{ourmodel} for both the case of  compressible and incompressible flows.

\paragraph{ \bf Compressible overdamped nematic.}
In this case, after eliminating the flow using Eq.~\eqref{fric_vel}, the linearized equations for the Fourier amplitudes of the fluctuations of $\Q$ about a uniform nematic state ($r>1$) are given by
%
\bea
&&\p_t Q_{xx} = -\f{1}{\gamma} \left[ 2 A (r-1) + K^{xx}(\a,\psi) q^2 + \kappa q^4\right] Q_{xx} \nn\\&&\hspace{0.5in}- \f{\lambda_2 \a}{2 \Gamma} S_0\sin 2 \psi ~q^2 Q_{xy}
\label{Qxx_comp}\;,\\
&&\p_t Q_{xy} = - \f{1}{\gamma} \left[ K^{xy}(\a,\psi) q^2 + \kappa q^2\right] Q_{xy}+ \f{\lambda_R \alpha }{2\Gamma}S_0 \sin 2\psi ~q^2 Q_{xx} \nn
\label{Qxy_comp}\;,
\eea
where  
\bea
&&K^{xx}(\a,\psi)=K+ \f{\lambda_1 \a}{2 \Gamma} + \f{\lambda_2 \a}{2\Gamma} S_0 \cos 2 \psi - \f{\lambda_3 \a}{4 \Gamma} S_0^2\;,\\
&&K^{xy}(\a,\psi)=K+\f{\lambda_1 \a}{ 2 \Gamma}+ \f{\lambda_R \a} {2 \Gamma}S_0 \cos 2 \psi\;,
\eea
 with $\psi$ the angle between the wave vector and the ordering direction.  When $\lambda_2=0$ and 
 $\lambda_1=\lambda_3=\lambda$, the above expressions reduce those given in Eqs. \ref{Ks} and \ref{Kt}
 of the main text.  The effective
stiffnesses $K^{xx}$ and $K^{xy}$ can always be driven negative at large extensile activities. Additionally, Eqs.~\eqref{Qxx_comp} and \eqref{Qxy_comp} decouple for pure bend and splay deformations corresponding  to $\psi=0$ (bend) and $\psi=\pi/2$ (splay).  Along these directions the instabilities of both director and order parameter magnitude have the same qualitative behavior as for the parameters used in the main text, with only quantitative differences. For general directions, the modes are coupled, but again the various instabilities and regimes are qualitatively unchanged.

\paragraph{\bf Incompressible overdamped nematic. }
When the flow is incompressible, it is not possible to obtain a single closed nonlinear PDE for $\Q$. One can still, however, explicitly eliminate the velocity in the  linearized equations for the fluctuations. The dynamics of the Fourier amplitude of fluctuations about the ordered state ($r>1$) is then  given by
%
\bea
&&\p_t Q_{xx} = -\f{1}{\gamma} \left[ 2A (r-1) + K^{xx}(\a,\psi) q^2 + \kappa q^4\right] Q_{xx} \nn\\
&&\hspace{0.5in}+ \f{\a}{2 \Gamma}\sin 2 \psi\left[\lambda_1 \cos2 \psi
 - \lambda_3S_0^2/2\right] q^2 Q_{xy} \;,
\label{Qxx_incomp}\\
&&\p_t Q_{xy} = -\f{1}{\gamma} \left[ K^{xy}(\a,\psi) q^2 + \kappa q^4 \right] Q_{xy} \nn\\
&&\hspace{0.5in}
+ \f{\a}{2 \Gamma} \sin 2 \psi\left[\lambda_1 \cos 2 \psi
 +\lambda_RS_0\right] q^2 Q_{xx} \;,
\label{Qxy_incomp}
\eea
where 
\bea
&&K^{xx}(\a,\psi)=K+\f{\a  \gamma}{2 \Gamma} \sin^2 2 \psi \l \lambda_1- \lambda_3 S_0^2/2\r\;,\\
&&K^{xy}(\a,\psi)= K + \f{\a \gamma }{2 \Gamma} \cos 2 \psi \l \lambda_1 \cos 2 \psi+ \lambda_R S_0\r\;.
\eea
%
%
%
Once again, the modes decouple for pure bend and splay fluctuations, with 
\bea
&&K^{xx}(\a,\psi=0)=K^{xx}(\a,\psi=\pi/2)=K\;,\\
&&K^{xy}(\a,\psi=0)= K + \f{\a \gamma \lambda_1}{2 \Gamma} \l 1+ \frac{\lambda_R S_0}{\lambda_1} \r\;,\\
&&K^{xy}(\a,\psi=\pi/2)= K + \f{\a \gamma \lambda_1}{2 \Gamma} \l 1- \frac{\lambda_R S_0}{\lambda_1} \r\;.
\eea
\noi
The fastest growing unstable mode is along the direction corresponding to pure bend for extensile systems and to pure splay for contractile ones, as shown in Fig.~\ref{polar_plot} for extensile systems, and describes the growth of director fluctuations. However, in incompressible systems the mode controlling the dynamics of fluctuations in the order parameter amplitude is stable along these special directions and
becomes unstable only in a narrow range of 
angles $\psi$ centered around $\psi=\pi/4$ \emph{and} in a narrow range of activity. 
 This difference may lead to the suppression of defect formation in frictional systems reported in Ref.~\cite{Amin:2016}, but verifying this
requires numerical investigations beyond 
the scope of the present work. 

\renewcommand{\thefigure}{B\arabic{figure}}

\setcounter{figure}{0}

\begin{figure}
\centering
{\includegraphics[scale=0.3]{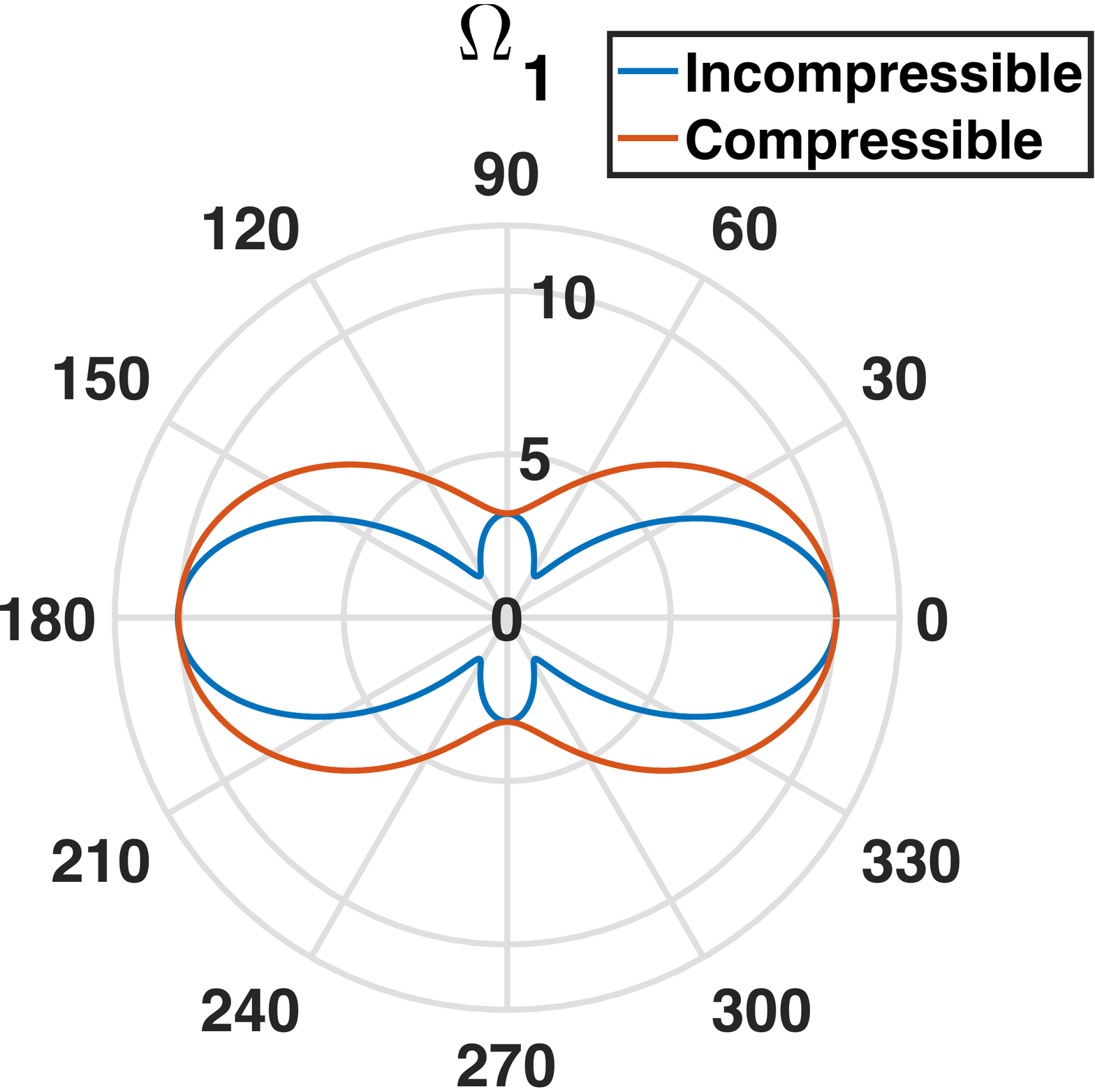}} 
\caption{Polar plot of the growth rate of the most unstable mode (this is the mode that for $\psi=0,\pi/2$ describes fluctuations in the director $\sim \Delta Q_{xy}$) as a function of the angle $\psi$ for fixed $q=0.7$,chosen such that all directions are unstable, and extensile systems. For both, incompressible (blue curve) and compressible (red curve) flows,  the  growth rate is largest for $\psi=0,\pi$, corresponding to bend fluctuations.   The parameter values used for the figure are $\lambda_1=1.5$, $\lambda_3=1.5$, $\lambda_R=1$, $\lambda_C=1$ and $\tilde{\a}=-20.0$, with units  chosen  as  in the main text.} 
\label{polar_plot}
\end{figure}

}

\end{appendices}

\bibliography{Revised_damped_active_nematic_31Aug16}
\bibliographystyle{rsc}

\end{document}